\documentclass[prl,twocolumn,showpacs,floats]{revtex4}
\usepackage{epsfig}
\usepackage{amsmath}
\newcommand{\be}{\begin{equation}}
\newcommand{\ee}{\end{equation}}
\newcommand{\bea}{\begin{eqnarray}}
\newcommand{\eea}{\end{eqnarray}}
\newcommand{\ba}{\begin{array}}
\newcommand{\ea}{\end{array}}
\newcommand{\nn}{\nonumber}
\newcommand{\parti}{\partial}

\newcommand{\al}{\alpha}
\newcommand{\sig}{\sigma}

\newcommand{\eps}{\epsilon}
\newcommand{\lam}{\lambda}
\begin{document}

\title{\bf Quantum Pumping and Nuclear Polarization in the Integer Quantum Hall Regime} 

\author{M. Blaauboer
}

\affiliation{Kavli Institute of NanoScience, Delft University of Technology,
Lorentzweg 1, 2628 CJ Delft, The Netherlands}
\date{\today}

\begin{abstract}
We study pumping of charge in a 2DEG in the quantum Hall regime
at filling factor $\nu = 2$ (2 spin-split levels of the lowest Landau level).
For pumping frequencies that match the Zeeman energy splitting, quantum pumping together with hyperfine interaction between electrons and nuclei induces transitions between the spin-split levels. These lead to a step-like change in the pumped current and to polarization of the nuclei. We present quantitative predictions for both. Our model provides the first quantitative tool to both control and measure the amount of local nuclear polarization in a 2DEG.
\end{abstract}

\maketitle

Charge pumping involves the generation of a d.c. current in the absence of an applied
bias voltage. The current is generated through application of two, or more,
time-varying signals with a phase difference $\phi$ between them. In recent years,
attention has focussed on pumping of charge in phase-coherent mesoscopic 
systems~\cite{brou98,alei98,swit99,wats03} such as e.g. quantum dots, small conducting islands embedded in
a semiconductor material~\cite{kouw97}. The pumped current in these systems depends on 
quantum-mechanical interference, hence the name ``quantum pumping'', and exhibits some fundamental properties, such as sinusoidal dependence on the
phase difference $\phi$ and linear dependence on the pumping 
frequency $\omega$ for $\omega \ll \tau_{d}^{-1}$, with $\tau_{d}$
the dwell time of the electrons in the system~\cite{brou98,swit99}. 

In this work, we investigate pumping of charge in a two-dimensional electron gas (2DEG) in the integer quantum Hall regime. Transport in this system occurs via edge states corresponding to quantized energy levels, the Landau levels (LLs)~\cite{bookIQH,datt95,been91}. Each LL is macroscopically degenerate, holding many electrons, and the number of occupied LLs is characterized by the filling factor $\nu\equiv n_e h/(eB)$, the number of electrons per flux quantum in the system [$n_e$ denotes the electron 
density and $B$ the applied magnetic field]. Our motivation to study quantum pumping in such a system comes from the idea that this way of transport might lead to interesting effects in the regime of spin-split LLs~\cite{simo00}. Since pumping involves exchange of energy quanta $\hbar \omega$, as described below, it can cause transitions between these LLs in combination with spin-dependent interactions in the system. At the same time, the pumped current can be used to monitor such transitions. This opens new possibilities to manipulate and measure spin-flip processes and nuclear polarization in a 2DEG.

\begin{figure}
\centerline{\epsfig{figure=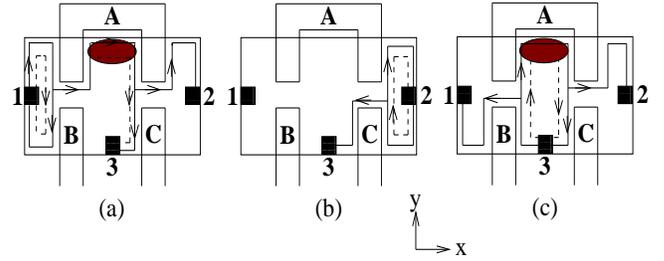,height=3.4cm,width=0.98\hsize}}
\caption[]{Schematic picture of a Hall bar at bulk filling factor $\nu=2$ (spin-split levels of the lowest Landau level). The split-gates A-B and A-C form two QPCs with conductance $G_{\rm AB}$,$G_{\rm AC} \leq e^2/h$ and are used for pumping by applying two time-varying voltages with a phase difference $\phi$ to B and C. The current leads 1-3 are all held at the same chemical potential $\mu$. Figs. (a)-(c) show the edge channels originating from, respectively, terminal 1, 2 and 3. Solid (dashed) lines correspond to the spin-up (spin-down) edge channel. In the shaded region spin-flip scattering can take place.
}
\label{fig:system}
\end{figure}

Using a Floquet scattering approach, we calculate the pumped current in the
Hall geometry of Fig.~\ref{fig:system}. Pumping is induced by applying ac voltages to the split-gates A-B and A-C on top of the bar, which results in
current flowing into contacts 1, 2, and 3. The gates form 2 quantum point contacts (QPCs) that transmit at most one edge channel. When the pumping frequency matches the Zeeman energy splitting, pumping together with hyperfine interaction between electrons and nuclei in the system causes scattering from the $\uparrow$-spin to the $\downarrow$-spin LL. This manifests itself as a sudden change of the pumped current and leads to polarization of the nuclei. For small pumping amplitudes, we calculate the change in current and the corresponding nuclear polarization analytically. The relationship between these two provides a tool to measure the amount of local nuclear polarization in the system. We conclude by discussing the validity of our model and comparing it to previous work on hyperfine interaction in quantum Hall systems.

Consider the Hall bar of Fig.~\ref{fig:system}. In the absence of the gates A, B, and C, the dynamics of the electrons is described by the Schr\"odinger equation $i\hbar \frac{\parti}{\parti t} \psi(x,y,t) = {\cal H} \psi(x,y,t)$ with
\be
{\cal H} = \frac{1}{2m^{*}} (i\hbar \vec{\nabla} + e \vec{A})^2 + V(y) +
\frac{1}{2} g^{*} \mu_B B S_z + \frac{1}{2} \sum_i A_i \vec{I}_i \cdot \vec{S}.
\label{eq:Ham}
\ee
Here $m^{*}$ denotes the effective mass, $\vec{A}$ the vector potential, $V(y)$ the transverse confining potential, $g^{*}$ the effective g-factor, and $\mu_B$ the Bohr magneton. The last term in Eq.(\ref{eq:Ham}) is the hyperfine interaction between the electron spin $\vec{S}$ and the nuclear spin $\vec{I}_i$, with $A_i$ the corresponding hyperfine interaction strength. If $V(y)$ is constant on the scale of the magnetic length $l_m \equiv \left( \frac{\hbar}{m^{*}\sqrt{\omega_0^2 + \omega_c^2}}\right)^{1/2}$, with $\omega_c\equiv |e|B/m^{*}$ the cyclotron frequency, it can be described to a good approximation by a harmonic potential $V(y) = \frac{1}{2} m^{*} \omega_0^2 y^2$~\cite{datt95,hale99}. Taking the Landau gauge $A_x=-By$, the Schr\"{o}dinger equation can be solved exactly in the mean-field approximation of the hyperfine interaction, with the result~\cite{laug98}
\bea
\psi_{\lam,\sig}(x,y,t) &=& C_{\lam}\, \chi_{\lam}(y)\, e^{i(kx - \frac{E_{\lam,\sig}}{\hbar}t)}, 
\label{eq:wavefunction} \\
\chi_{\lam}(y) & = & e^{\frac{1}{2 l_m^2}(y+\frac{\omega_c^2}{\omega_0^2+\omega_c^2} y_k)^2}
\left( \frac{\parti}{\parti y} \right)^{\lam} e^{-\frac{1}{l_m^2}(y+\frac{\omega_c^2}{\omega_0^2+\omega_c^2} y_k)^2} \\
E_{\lam,\sig} & = & (\lam + \frac{1}{2}) \hbar \sqrt{\omega_0^2 + \omega_c^2} + \frac{1}{2} m^{*} \frac{\omega_{0}^2 \omega_c^2}{\omega_{0}^2 + \omega_c^2}
y_{k}^2 \nn \\
&& + \frac{1}{2} g^{*} \mu_B B \sig + \frac{1}{2} A \langle I_z \rangle \sig,\ \ \ \lam = 0,1,2,...
\label{eq:LLs}
\eea
Here $C_{\lam}$ denotes a normalization constant, $y_{k} \equiv \hbar k/eB$, $\sig = 1\, (-1)$ for electrons with spin-up (down), and $A\equiv \sum_i A_i$. $\langle I_z \rangle$ is the average nuclear polarization and $E_{\lam,\sig}$ represent the energy levels (LLs) measured from the lowest band imposed by the confinement in the $z$-direction. The electrons travel near the edges of the 2DEG along one-dimensional (1D) channels that are formed at the intersection of the Fermi energy $E_{F}$ with the LLs (\ref{eq:LLs})~\cite{been91}. 

\begin{figure}
\centerline{\epsfig{figure=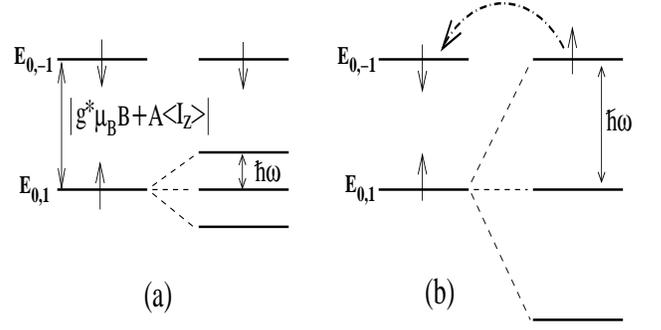,height=4.3cm,width=0.95\hsize}}
\caption[]{Energy level diagrams for (a) $\hbar \omega \ll | g^{*} \mu_B B + A \langle I_z \rangle|$ and (b) $\hbar \omega \sim |g^{*} \mu_B B + A \langle I_z \rangle|$. See text for discussion.
}
\label{fig:levels}
\end{figure}

In order to study quantum pumping in this system, we model the QPCs in Fig.~\ref{fig:system} as oscillating $\delta$-function potentials, $V_{AB}(x,t)= [\bar{V}_{AB} + 2\, \delta V_{AB} \cos (\omega t + \phi_{AB})]\, \delta (x + L/2)$ and $V_{AC}(x,t)= [\bar{V}_{AC} + 2\, \delta V_{AC} \cos (\omega t + \phi_{AC})]\, \delta (x - L/2)$ . We then use Floquet theory~\cite{Floquet} to calculate the pumped current flowing along the edge channels. This approach has recently been employed to study 1D mesoscopic quantum pumps~\cite{zhu02,mosk02} and describes pumping in terms of gain or loss of energy quanta $\hbar \omega$ at the two oscillating barriers. Following Moskalets and B\"{u}ttiker~\cite{mosk02}, the pumped current into lead $\al$, with $\al=1,2,3$, is given by
\bea
I_{\al} & = & \frac{e}{h} \int_{0}^{\infty} dE \sum_{\beta} \sum_{E_n >0}
|t_{\al \beta}(E_n,E)|^2 \left( f(E) - f(E_n) \right),
\label{eq:currmosk}
\eea
where $f(E)=(1+ {\rm exp}[(E-\mu)/(k_B T)])^{-1}$ is the Fermi distribution, with $\mu$ the chemical potential in the leads, and $t_{\al \beta}(E_{n},E)$ denotes the Floquet scattering amplitude for a particle originating with energy $E$ in lead $\beta$ to gain $n$ energy quanta $\hbar \omega$ and be scattered with energy $E_n \equiv E + n \hbar \omega$ into lead $\al$. For two oscillating barriers, these scattering amplitudes are calculated by matching wave functions (see Eq.~(\ref{eq:wavefunction})) $\psi(x,t) = \sum_{n=-\infty}^{\infty} (a_{n} e^{i k_n x} + b_{n} e^{-i k_n x}) e^{-i(E_{0,1}+n \hbar \omega)t/\hbar}$ at each of the two barriers~\cite{mosk02}. For details of this calculation we refer to Ref.~\cite{blaa03}. Here we restrict ourselves to the situation of weak pumping, defined by $q_{AB(AC)} \equiv \delta V_{AB(AC)} m^{*}/\hbar^2 \ll k_0$, with $k_n = \sqrt{\frac{2m^{*}}{\hbar^2} \frac{\omega_0^2+\omega_c^2}{\omega_0^2}} \left[ E_{F} + n\hbar \omega - \frac{1}{2} \hbar \sqrt{\omega_0^2 + \omega_c^2} - \frac{1}{2} g^{*} \mu_B B - \right.$ $\left. \frac{1}{2} A \langle I_z \rangle \right]^{1/2}$. In this case it is sufficient to take only the two lowest sidebands $n=0, \pm 1$ into account~\cite{blaa03} and the pumped current can be obtained analytically. Before analyzing this current, let us take a closer look at the energy levels in the region between the QPCs. These are drawn schematically in Fig.~\ref{fig:levels} for
pumping frequencies much less than [Fig.~\ref{fig:levels}(a)] and approximately equal to [Fig.~\ref{fig:levels}(b)] the effective Zeeman splitting.

In each picture the left side represents the energy levels corresponding to the spin-up and spin-down edge channels {\it before} scattering at a QPC (describing the situation in, e.g., the upper left corner of Fig.~\ref{fig:system}(a)), while the right half represents the situation {\it after} scattering at a QPC. In the latter, electrons in the spin-up channel either remain at energy $E_{0,1}$ or are scattered into sidebands with energies $E_{0,1} \pm \hbar \omega$. The $\downarrow$-spin electrons, which are not in touch with the oscillating barrier (since the QPC transmits at most one, the $\uparrow$-spin, edge channel), all remain at $E_{0,-1}$. This lifting of degeneracy of the $\uparrow$-spin level has an interesting consequence: for appropriate level splitting $\hbar \omega$, it can combine with hyperfine interaction to enable scattering from the $\uparrow$-spin to the $\downarrow$-spin band. If $\hbar \omega$ is less than the effective Zeeman splitting $|g^{*} \mu_B B + A \langle I_z \rangle|$, as in Fig.~\ref{fig:levels}(a), this is not possible, since the $\downarrow$-spin level is too high in energy. If, however, $\hbar \omega$ is equal to $|g^{*} \mu_B B + A \langle I_z \rangle|$ (Fig.~\ref{fig:levels}(b)), electrons in the ($E_{0,1}+\hbar \omega$)-level can flip-flop scatter into the $\downarrow$-spin level. As a result, part of the spin-up electrons that would normally, i.e. in the absence of spin-flip scattering, flow into lead 2, now end up as spin-down electrons in lead 3~\cite{flipflop}. One would expect this to lead to a visible change in $I_3$, the current pumped into contact 3.

We now calculate $I_1$, $I_2$ and $I_3$ from Eq.~(\ref{eq:currmosk}). Assuming equal barriers ($q_{AB}=q_{AC}\equiv q$ and $\bar{V}_{AB} = \bar{V}_{AC} \equiv \hbar^2 p/m^{*}$), low temperatures ($k_B T \ll \hbar \omega$) and no spin-flip scattering in the region between the QPCs we obtain in the adiabatic limit,
\begin{subequations}
\bea  
I_{1}^{\rm no\ sf} & = & 0 \label{eq:currents11} \\
I_{2}^{\rm no\ sf} & = & - \frac{2 e \omega}{\pi} \, \frac{k_0^3 p q^2 
(k_0^2 + p^2 - 2 q^2) \sin(4\eps k_0 L + \phi)}{[(k_0^2 + p^2)^2 + 4q^2(k_0^2-p^2)+4 q^4]^2}  \\ 
I_{3}^{\rm no\ sf} & = & \frac{2 e \omega}{\pi} \, \frac{k_0^3 p q^2 
(k_0^2 + p^2 - 2 q^2) \sin(4\eps k_0 L + \phi)}{[(k_0^2 + p^2)^2 + 4q^2(k_0^2-p^2)+4 q^4]^2}. 
\eea
\label{eq:currents1}
\end{subequations}
Here $\eps \equiv \hbar \omega/ (4[E_F - \frac{1}{2} \hbar \sqrt{\omega_0^2 + \omega_c^2} - \frac{1}{2} g^{*} \mu_B B - \frac{1}{2} A \langle I_z \rangle])$~\cite{eps} and $\phi \equiv \phi_{AC} - \phi_{AB}$. In the opposite limit of full spin-flip scattering, i.e. when each electron in the $\uparrow$-spin channel in the region between the two QPCs is scattered into the $\downarrow$-spin channel, the result is
\begin{subequations}
\bea
I_{1}^{\rm sf} & = & 0 \\
I_{2}^{\rm sf} & = & - \frac{e \omega}{\pi} \, k_0^4 q^2 [ 
(k_0^2 + p^2) + (k_0^2 + p^2 - 2 q^2) \cos (4\eps k_0 L + \phi)  
\nn \\
&& + \frac{p}{k_0} (k_0^2 + p^2 - 2 q^2) \sin(4\eps k_0 L + \phi)]/ \nn \\
&& \ \  [(k_0^2 + p^2)^2 + 4 q^2 (k_0^2 - p^2)+ 4 q^4]^2 \\
I_{3}^{\rm sf} & = & \frac{e \omega}{\pi} \, k_0^4 q^2 [ 
(k_0^2 + p^2) + (k_0^2 + p^2 - 2 q^2) \cos (4\eps k_0 L + \phi)  
\nn \\
&& + \frac{p}{k_0} (k_0^2 + p^2 - 2 q^2) \sin(4\eps k_0 L + \phi)]/ \nn \\
&& \ \  [(k_0^2 + p^2)^2 + 4q^2(k_0^2-p^2)+4 q^4]^2
\eea
\label{eq:currents2}
\end{subequations}
Note that in both cases the total current $I_1+I_2+I_3=0$. Eqns.~(\ref{eq:currents1}) and (\ref{eq:currents2}) are proportional to the pumping frequency $\omega$ and amplitude $q^2$, which is characteristic of adiabatically pumped currents~\cite{brou98,alei98}. The currents in general do not vanish if the phase difference $\phi=0$, though. This is due to the 3-terminal geometry considered here: for the corresponding 2-terminal Hall bar one obtains $I_{1,2} \sim q^2 \sin \phi \sin (2 \eps k_0 L) \stackrel{\eps k_0 L \ll 1}{\sim} q^2 \omega \sin \phi$. In the presence of a third (symmetry-breaking) terminal, current can be pumped even if the two barriers oscillate in phase and with the same amplitude.

\begin{figure}
\centerline{\epsfig{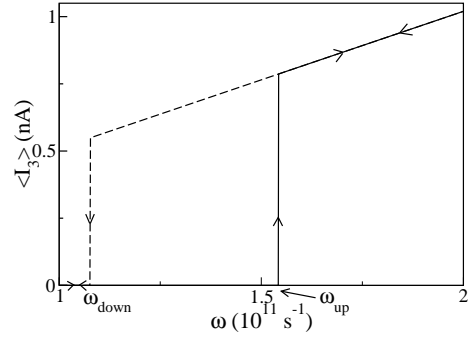}}
\caption[]{The averaged pumped current $I_3$ (Eqs.~(\ref{eq:currents33}) and (\ref{eq:currents43})) for a linear sweep of the pumping frequency $\omega$, see the text for explanation. Parameters used are $B=4\, T$, $\phi=\pi/2$, $q^2/k_0^2=0.1$, $g^{*}=-0.44$, $m^{*}=0.61\, 10^{-31}$ kg, $A=90\, \mu$eV~\cite{merk02}, $t_{\rm sweep} = 10^{-5}$ s, $t_{\rm dwell} = 10^{-5}$ s, $v = 10^{16}$ s$^{-2}$ and $N_{n}=10^7$.
}
\label{fig:I3}
\end{figure}

Eqs.~(\ref{eq:currents1}) and (\ref{eq:currents2}) simplify in case of nearly-open QPCs, corresponding to low barriers with $p \ll k_0$ and $2q<p$~\cite{open}. The pumped currents then become, up to O($\frac{p}{k_0}$),
\begin{subequations}
\bea  
I_{1}^{\rm no\ sf} & = & 0 \label{eq:currents31} \\
I_{2}^{\rm no\ sf} & = & -2 \frac{e \omega}{\pi}\, \frac{q^2}{k_0^2}\, \frac{p}{k_0}\, \sin(4\eps k_0 L + \phi)  \\
I_{3}^{\rm no\ sf} & = & 2 \frac{e \omega}{\pi}\, \frac{q^2}{k_0^2}\, \frac{p}{k_0}\, \sin(4\eps k_0 L + \phi),
\label{eq:currents33}
\eea
\label{eq:currents3}
\end{subequations}
and
\begin{subequations}
\bea  
I_{1}^{\rm sf} & = & 0 \\
I_{2}^{\rm sf} & = & - \frac{e \omega}{\pi}\, \frac{q^2}{k_0^2}\, \left(
1 + \cos(4\eps k_0 L + \phi) + \frac{p}{k_0} \sin(4\eps k_0 L + \phi) 
\right) \\
I_{3}^{\rm sf} & = & \frac{e \omega}{\pi}\, \frac{q^2}{k_0^2}\, \left(
1 + \cos(4\eps k_0 L + \phi) + \frac{p}{k_0} \sin(4\eps k_0 L + \phi) 
\right) \label{eq:currents43}.
\eea
\label{eq:currents4}
\end{subequations}
Note the dramatic change in $\langle I_3 \rangle$ ($I_3$ averaged over $k_0L$),  from 0 for no spin flip scattering to the finite value $\frac{e \omega}{\pi}\, \frac{q^2}{k_0^2}$ at the onset of full spin-flip scattering. This change in $\langle I_3 \rangle$ can be used to detect the amount of nuclear polarization in this region~\cite{I2}. In order to illustrate this, Fig.~\ref{fig:I3} shows the behavior of $\langle I_3 \rangle$ if the pumping frequency is linearly swept up during a time $t_{\rm sweep}$ as $\omega(t) = \omega(0) + v t$ [with $v$ the sweep rate], then kept at the value $\omega(t_{\rm sweep})$ during a time $t_{\rm dwell}$, and subsequently swept down again as $\omega(t) = \omega(t_{\rm sweep}) - v (t-t_{\rm sweep} - t_{\rm dwell})$ for $t_{\rm sweep} + t_{\rm dwell} \leq t \leq 2 t_{\rm sweep} + t_{\rm dwell}$.

Starting from $\omega(0)=1 \cdot 10^{11} s^{-1}$, the pumped current into lead 3 is first zero, since for low frequencies no spin-flip scattering can occur (Fig.~\ref{fig:levels}(a)). Then, assuming the nuclei to be initially unpolarized, at frequency $\omega_{\rm up} = |g^{*}| \mu_B B/\hbar$ electrons begin to scatter from the $\uparrow$-spin band to the $\downarrow$-spin band (Fig.~\ref{fig:levels}(b)), leading to a jump in $I_3$ to the value $\frac{e \omega}{\pi}\, \frac{q^2}{k_0^2}$ (assuming maximal spin-flip scattering, see the discussion below). Sweeping the frequency further up continues to increase the amount of nuclear polarization through spin-flip scattering~\cite{phonons}. Upon reversing the sweep direction, the current decreases while the nuclear polarization still increases, until at frequency $\omega_{\rm down}$ the highest energy level of the $\uparrow$-spin falls below that of the $\downarrow$-spin, reducing spin-flip scattering and the current to zero. Because the nuclei are now polarized with average polarization $\langle I_z \rangle (t) > 0$, $\omega_{\rm down}$ is smaller than $\omega_{\rm up}$ and a hysteresis loop is formed, see Fig.~\ref{fig:I3}. For Ga and As nuclei with $I=3/2$, $\langle I_z \rangle (t)$ = 3 $\times$ the number of flipped spins per unit of time $\times$ $(t-t_{\rm up})$/ total number of nuclear spins, where $t_{\rm up}$ corresponds to the time at which $\omega_{\rm up}$ is reached, so $t_{\rm up} = (|g^{*}| \mu_B B - \hbar \omega(0))/(\hbar v)$. In case of nearly-open QPCs this yields
\be
\langle I_z \rangle (t) = 3\, \frac{\omega_{\rm av}}{\pi N_n}\, \frac{q^2}{k_0^2}\, (t-t_{\rm up})
\label{eq:nuclpol}
\ee
for $t_{\rm up} \leq t \leq t_{\rm down}$, with $N_n$ the total number of nuclei in the scattering region, $\langle I_z \rangle \leq 3/2$ $\forall t$ and $\omega_{\rm av}$ the average frequency during $t - t_{\rm up}$. The difference $\Delta \omega \equiv \omega_{\rm up} - \omega_{\rm down}$ is given by
\bea
\Delta \omega & = & \frac{M}{\hbar (M-\hbar v)} \left[\, 2 |g^{*}| 
\mu_B B - 2 \hbar \omega(0) - \right. \nn \\ 
&& \left. \hspace*{2cm} \hbar v (2t_{\rm sweep} + t_{\rm dwell}) \right],
\label{eq:freqdiff}
\eea
with $M \equiv 3\,\frac{\omega_{\rm av}}{\pi N_n}\, \frac{q^2}{k_0^2} A$ and provided $M(t_{\rm down} - t_{\rm up}) \leq |g^{*}| \mu_B B$. $\Delta \omega$ is thus a direct measure of $N_n$. This opens the possibility to both {\it control} (through $t_{\rm sweep}$, $t_{\rm dwell}$ and $v$) and accurately {\it estimate} (via Eq.~(\ref{eq:nuclpol})) the amount of nuclear polarization by changing the pumping frequency $\omega$ and measuring $I_3$. The sharpness of the steps in Fig.~\ref{fig:I3} is due to the mean field approximation of the nuclear polarization $\langle I_z \rangle$ and is reduced if one goes beyond that approximation.

Finally, let us discuss spin-flip times due to hyperfine interaction and previous work on nuclear spin effects in 2DEGs. We have recently calculated the spin-flip time $t_{\rm sf}$ due to hyperfine interaction in a quantum channel in the presence of a magnetic field~\cite{blaa04}. Using this result to calculate the spin-flip time $t_{\downarrow \uparrow}$ from the $\uparrow$-channel to the $\downarrow$-channel we obtain for typical GaAs parameters $t_{\downarrow \uparrow} <$ 10 ns, while the reverse scattering time $t_{\uparrow \downarrow} \gg t_{\downarrow \uparrow}$ due to full occupancy of the $\uparrow$-spin edge channel. The travel time of an electron in the region between the two QPCs is $t_{\rm travel} \sim L/v_{\rm ch} \sim 5 \cdot 10^{-8}$ s for $L = 50 \mu$m and $v_{\rm ch} = \frac{\hbar k}{m^{*}}\, \frac{\omega_0^2}{\omega_c^2 + \omega_0^2}$ with $\omega_0 \approx 0.08\, \omega_c$~\cite{hale99}. Since $t_{\downarrow \uparrow} < t_{\rm travel} \ll t_{\uparrow \downarrow}$, the probability that an electron with spin-$\uparrow$ will scatter to the $\downarrow$-spin channel is very large and we can to good approximation assume full spin-flip scattering. We then find that $I_{1}, I_{2}, I_3 \sim 0.1-1$ nA and for timescales $t_{\rm dwell} \sim 10^{-5}$ s shifts of $\sim 30 \%$ of the Zeeman energy gap, which is well within reach of observation.

Nuclear spin effects at the edges of a 2DEG have been studied before in the past decade~\cite{wald94,dixo97,mach03}. In Ref.~\cite{dixo97}, Dixon {\it et al.} examine a similar 3-terminal geometry at $\nu=2$. They measure $I-V$ curves, which show hysteretic behavior and provide strong evidence for dynamic nuclear polarization through hyperfine interaction at the edges. The new element in our work here is the use of quantum pumping instead of applying a bias voltage to generate a current. This allows, as far as we know for the first time, to quantitatively predict the current flow between spin-split edge states and the amount of local nuclear polarization. Our simple model is based on a number of assumptions, for example neglect of exchange interactions and correlation effects between the edge channels. A rigorous theory should take these into account. However, based on experimental results in similar geometries \cite{dixo97}, we believe that the non-interacting particle picture presented here provides a valid first-order description of the basic physics of quantum pumping in spin-split edge states.

In conclusion, quantum pumping is a useful tool to study and probe local dynamic nuclear polarization at the edge of a 2DEG in the quantum Hall regime. Our model provides the first quantitative technique to accurately manipulate and detect local nuclear spin polarization. This is a topic of great recent interest, not in the least in view of possibilities to develop spin qubits in solid-state systems~\cite{qubits}.

Stimulating discussions with M. B\"{u}ttiker and M. Heiblum are gratefully acknowledged. This work has been supported by the Stichting voor Fundamenteel Onderzoek der Materie (FOM), and by the EU's Human Potential Research Network under contract No. 
HPRN-CT-2002-00309 (``QUACS'').


\begin{thebibliography}{99}
\bibitem{flipflop}
Spin-flip scattering also occurs in the regions to the left and
the right of both QPCs. There, however, even though the electrons flip their spin, they end up in the same lead as they would have otherwise, and the amount of spin flips would only be detectable in a spin-sensitive current measurement. We do not consider this possibility here.
\bibitem{eps}
In deriving Eqs.~(\ref{eq:currents1}) and (\ref{eq:currents2})  we have assumed that $\eps \ll 1$, so that $k_{\pm 1} \approx k_0$ as a prefactor~\cite{blaa03}, and that the electron gas below the two split gates is fully depleted, see the discussion in section V of Ref.~\cite{dixo97}.
\bibitem{open}
so that $G_{AB}, G_{AC} \leq \frac{e^2}{h}$ during the entire pumping cycle.
\end{thebibliography}
\end{document}